\documentclass[12pt]{article}
\usepackage{amsmath,amssymb}

\newcommand{\bea}{\begin{eqnarray}}
\newcommand{\eea}{\end{eqnarray}}
\newcommand{\be}{\begin{equation}}
\newcommand{\ee}{\end{equation}}
\newcommand{\vs}[1]{\vspace{#1 mm}}

\newcommand{\dsl}{\pa \kern-0.5em /}

\newcommand{\pa}{\partial}

\newcommand{\nn}{\nonumber\\}

\begin{document}
\topmargin 0mm
\oddsidemargin 0mm

\begin{flushright}

USTC-ICTS/PCFT-23-34\\

\end{flushright}

\vspace{2mm}

\begin{center}

{\Large \bf The non-perturbative stringy interaction between NS-brane \& Dp brane \\}

\vs{10}

{\large J. X. Lu and Nan Zhang}

\vspace{4mm}

{\em
Interdisciplinary Center for Theoretical Study\\
 University of Science and Technology of China, Hefei, Anhui
 230026, China\\
 \medskip
 Peng Huanwu Center for Fundamental Theory, Hefei, Anhui 230026, China\\ 
 %\vs{4}
}

\end{center}

\vs{10}

\begin{abstract}
To our best knowledge, the leading non-perturbative stringy interaction between an NS brane and a Dp brane remains unknown. We here present the non-perturbative stringy amplitudes for a system of an F-string and a Dp brane and a system of an NS 5 brane and a Dp brane for $0 \le p \le 6$. In either case,  the F or NS5 and the Dp are placed parallel at a separation.   We obtain the respective amplitudes, starting from the amplitude for a system of a D1 and a D3 for the former and that for a system of a D5 and a D3 system for the latter, based on the IIB S-duality and various T-dualities plus the consistency of both, along with the respective known long-range amplitudes.  We would like to point out that the amplitude for the D1/D3 or D3/D5 computed from the usual D-brane technique does not take into consideration of the non-perturbative contribution due to the exchange of virtual closed D-string emitted by the D3.  As such the resulting amplitudes obtained from this one via the S-duality and followed by various T-dualities are not consistent with the IIB S-duality. We resolve this issue and  obtain the corresponding consistent amplitudes. The implications of so obtained amplitudes are also discussed.  
 \end{abstract}

\newpage

\section{Introduction}
In general, one can deduce the interaction nature (attractive, repulsive or zero) of one Dp and the other Dp$'$ with $p - p' = 2\, n$ (here $n = 0, 1, 2, 3$), placed parallel at a separation, when they each carries a constant worldvolume magnetic  flux. This is due to that the magnetic flux represents the codimension-2 D-branes within the original one\footnote{The corresponding system gives the so-called (D(p - 2), Dp) or (D(p$'$ - 2), Dp$'$) non-threshold bound state when $p \ge p' \ge 2$ \cite{Witten:1995im, Breckenridge:1996tt, Costa:1996zd, Gava:1997jt, Di Vecchia:1997pr, Sheikh-Jabbari:1997qke}.} and we know the interaction nature of a Dp and a Dp$'$  in a similar setting when they each carries no flux at all\footnote{When $p - p' =0$ or $4$, there is no interaction between the two, when $p - p' = 2$, the interaction is attractive while $p - p' = 6$, the interaction is repulsive.}. Sometimes T-dualities are needed for such an understanding. 

However, the story is different when a constant worldvolume electric flux, which represents the fundamental strings (for short F-strings) delocalized along the directions transverse to the flux direction within the Dp brane\footnote{Such a system is usually called non-threshold bound state (F, Dp)  \cite{Schwarz:1995dk, Lu:1999uca,DiVecchia:1999uf}. The existence of such bound states  was shown in \cite{Witten:1995im}. Related works using mixed boundary conditions for these bound states were discussed in \cite{Arfaei:1997hb, Sheikh-Jabbari:1997qke}.},  is present.  This is due to that we don't know, to our best knowledge, the non-perturbative stringy interaction between an F-string and a Dp brane in general except for the large-separation one which can be computed using effective field theories, for example, see \cite{Ouyang:2014bha}. 

D-branes are a sort of non-perturbative with respect to the F-string.  For a weak string coupling, D-branes can be viewed as rigid at least to the scale up to string one or even smaller, for example, see the discussion given in \cite{Lu:2007kv}.   How to compute the non-perturbative stringy interaction between an F-string or an NS5 brane and a Dp brane appears to be a difficult and yet unsolved problem to our best knowledge though the corresponding long-range one was given a while ago, for example, see  \cite{Ouyang:2014bha}.  The computed long-range interaction per unit Dp-brane worldvolume with $0 \le p \le 6$ in coordinate space can be given as\footnote{Our conventions are that the positive sign of the interaction corresponds to an attractive one while the negative sign corresponds to a repulsive one.}
\be\label{long-rangeF/Dp}
U_{\rm{F/Dp}} \, (y) = \frac{\eta_{p} \, V_{2}\, g_{s}\, \pi \left(2 \pi \sqrt{\alpha'}\right)^{5 - p}}{(7 - p)\, y^{7 - p}\, \Omega_{8 -p}},
\ee
where $V_{2}$ is the volume of the F-string worldvolume,  $g_{s}$ the string coupling constant and $y$ the separation between the F-string and the Dp with $y^{2} = y^{2}_{p + 1} + \cdots y^{2}_{9}$.  This can be obtained from that given in \cite{Ouyang:2014bha} in momentum space using the following     
\be
\int \frac{d^{\perp} k_{\perp}}{(2\pi)^{\perp}} \frac{e^{- i {\bf k}_{\perp} \cdot {\bf y}}} {k^2_{\perp}} = \frac{1}{(7 - p) \Omega_{8 - p} \, y^{7 - p}},
\ee
where $\perp$ denotes the spatial dimensionality of the space transverse to the Dp brane under consideration which is $9 - p$. The parameter $\eta_{p} = 0$ when $p = 0$ while $\eta_{p} = 1$ for $p > 0$.  In the above, $\Omega_{n} = 2 \pi^{(n + 1)/2} /\Gamma ((n + 1)/2)$ the volume of unit n-sphere. The corresponding non-perturbative stringy amplitude, which we are going to obtain, is expected to agree with (\ref{long-rangeF/Dp}) at large separation though the two approaches are rather different.  Nevertheless, this agreement serves as a non-trivial consistent check of the non-perturbative stringy amplitude so obtained.  

In a similar fashion, we have the long-range interaction between an NS5 brane and a Dp brane with $0 \le p \le 5$ as
\be\label{long-rangeDp/NS5}
U_{\rm{NS5/Dp}} \, (y) = \frac{V_{p + 1} \, (2\pi \sqrt{\alpha'})^{1 - p}}{4 \pi\, g_{s} y^{2}},
\ee
where $V_{p + 1}$ is the volume of Dp brane worldvolume and $y$ the brane separation between the NS5 and the Dp.  There exists, however, no long-range interaction between an NS5 and a D6, placed parallel at separation, as given in \cite{Ouyang:2014bha}. This is consistent with the vanishing result of $U_{\rm{F/D0}} \, (y)$ since the two are related to each other by the 10D  Hodge duality.  

One can check easily that either the interaction (\ref{long-rangeF/Dp}) or (\ref{long-rangeDp/NS5}) or both are consistent with the IIB S-duality for odd $p$. This is also a point which we will use to check the relevant consistency for the corresponding non-perturbative stringy amplitudes. We would like also to stress that the leading interaction amplitudes (\ref{long-rangeF/Dp}) and (\ref{long-rangeDp/NS5}) are good for any string coupling $g_{s}$. 

In using the usual D-brane technique to compute the closed string tree cylinder amplitude between a Dp and a Dp$'$ with $p - p' = 2n$ with $n = 0, 1, 2, 3$, one only considers the contribution from the exchange of  virtual closed F-strings emitted by the D branes involved. This misses the important non-perturbative contribution from the exchange of virtual closed D-branes. One clear example is the D3 brane which can emit not only the virtual closed F-strings but also the virtual closed D-strings.  Consider a system of a Dp and a D3 with odd p, the D3 emitting a virtual closed F-string, propagating for some time and finally absorbed by the Dp gives rise to the usual closed string tree cylinder amplitude for the system.  The non-perturbative contribution from the D3 emitting a virtual closed D-string, propagating for some time and finally absorbed by the Dp has not been considered so far.  Precisely because of this, when we apply the IIB S-duality  to the usual closed string tree cylinder amplitude and follow by various T-dualities, the resulting amplitudes are not consistent with the IIB S-duality.  Consistency of the IIB S-duality along with various T-dualities forces upon us to include the corresponding non-perturbative contribution to the respective amplitude which we will demonstrate in this paper.  

Applying T-duality to the D3, the emitted virtual closed D-string will be changed to other virtual closed D-brane with possible different dimensionality. The tree cylinder contribution from the exchange of the virtual closed D-string becomes a one loop annulus one due to a virtual open D-string with its two ends connecting the D3 and the Dp if we make a Jacobi transformation to the original one. Applying T-duality to this open D-string one loop annulus contribution will change this open D-string to the other open D-brane \cite{Strominger:1995ac, Townsend:1996em} also with a different dimensionality.  So knowing the corresponding contribution to the underlying amplitude in either case will provide important non-perturbative information about the dynamics of the closed D-brane or the open D-brane which may not be able to obtain otherwise. 

This paper is arranged as follows. In section 2, we first discuss the case of an F-string and a D0 brane. For this case, the vanishing of the long-range interaction hints already that it vanishes in general. We will use the interaction amplitude for a D0 and a Dp carrying a constant electric flux, representing the presence of F-strings,  to deduce the result. Here we take $p = 2, 4, 6$, respectively. In section 3, we first compute the closed string cylinder amplitude using the usual D-brane technique for a D3 and a D1, placed parallel at a separation, and then use the IIB S-duality to obtain the corresponding one for a D3 and an F-string. Once this amplitude is obtained, we can use T-duality along directions either transverse or parallel (but not along the F-string) to D3  to obtain the corresponding amplitude for the F/Dp system with\footnote{We limit $p \le 6$ such that the large separation limit of the stringy amplitude can be compared with (\ref{long-rangeF/Dp}).} $ 1 \le p \le 6$. In section 4, we compute the amplitude for a D3 and a D5, again placed parallel at a separation. We once again use the IIB S-duality to obtain the corresponding amplitude for the D3/NS5. Once we have this, we can use T-duality along a longitudinal direction of NS5 at a time to obtain the respective amplitude for Dp/NS5 with $0 \le p \le 5$.  The amplitude for D6/NS5 vanishes which can be deduced from the one for F/D0 obtained in section 2 using the 10D  Hodge duality. This is also consistent with its long-range result.  However, there is an issue for the amplitudes so obtained in Type IIB string theory if we naively follow the procedures given in the previous two sections. Some of these IIB amplitudes appear not consistent with the IIB S-duality.  We in section 5 propose to resolve this issue and obtain the corresponding ones being consistent with the IIB S-duality. At the same time, we understand what is missing behind.  We conclude this paper in section 6.

\section{The F/D0 case}
 In general, if the long-range interaction vanishes due to its coefficient while being independent of the separation, this usually implies that the leading stringy one vanishes, too. In the following, we will make consistent checks of this by considering the closed string tree cylinder amplitude between a D0 and a (F, Dp), placed parallel at a separation $y$, with $p = 2, 4, 6$, respectively.  For this, we take the electric flux on the Dp as
\be\label{dpflux}
 \hat F_{p} =\left( \begin{array}{cccc}
0 & \hat f & 0& \ldots\\
- \hat f&0&0&\ldots\\
0&0&0&\ldots\\
 \vdots&\vdots&\vdots&\ddots
 \end{array}\right)_{(1 + p)\times (1 + p)},
\ee
where the dimensionless flux $\hat F_{\alpha\beta} = 2 \pi \alpha' F_{\alpha\beta}$ with $F_{\alpha\beta}$ the usual electromagnetic flux,  $2 \pi \alpha'$ the inverse of fundamental string tension and $\alpha, \beta = 0, 1, \cdots, p$.  Note that the interaction amplitude has an SO(p) rotation symmetry and as such we can choose the electric flux along `1' direction as given above without loss of generality. 
 The cylinder amplitude per unit p-brane worldvolume was computed recently in \cite{Lu:2023jxe}, and is respectively as 
\be\label{20cA}
\Gamma_{2/0} = \frac{2 V_{1}\sqrt{1 - {\hat f}^{2}}}{(8\pi^{2} \alpha')^{1/2}} \int_{0}^{\infty} \frac{d t}{t^{7/2}} \, e^{- \frac{y^{2}}{2\pi \alpha' t}} \prod_{n = 1}^{\infty} \frac{(1 +  |z|^{4n})^{4}}{(1 - |z|^{2n})^{6} (1 + |z|^{2n})^{2}},
\ee
for (F, D2)/D0, 
\be\label{40cA}
\Gamma_{4/0} =  \frac{ V_{1}\,  \sqrt{1 - {\hat f}^{2}} \,  \cos^{2}\pi \nu_{0}}{(8\pi^{2} \alpha')^{1/2}} \int_{0}^{\infty} \frac{d t}{t^{5/2}} \,  e^{- \frac{y^{2}}{2\pi \alpha' t}},
\ee
for (F, D4)/D0 with $\nu_{0} = 1/2$, and 
\be\label{60cA}
\Gamma_{6/0} = - \frac{ V_{1}\sqrt{1 - {\hat f}^{2}}}{2 (8\pi^{2} \alpha')^{1/2}} \int_{0}^{\infty} \frac{d t}{t^{3/2}} \, e^{- \frac{y^{2}}{2\pi \alpha' t}} \prod_{n = 1}^{\infty} \frac{(1 + |z|^{4n})^{4}}{(1 - |z|^{2n})^{2} (1 + |z|^{2n})^{6}},
\ee
for (F, D6)/D0. In the above, $V_{1}$ is the volume of D0 worldvolume and $|z| = e^{- \pi t}$.

 In the above for (F, D4)/D0, the amplitude with $\nu_{0} = 1/2$  actually vanishes. Since there is no interaction between  a D0 and a D4, this zero amplitude must imply that the interaction between  a D0 and an F-string vanishes, too,  a consistent check of the expected.  For (F, D2)/D0, this amplitude, apart from the overall factor $\sqrt{1 - {\hat f}^{2}}$ which is due to the normalization of the D2 boundary state in the presence of an electric flux., is exactly the same as that between a pure D2 and a D0.  This hints also the vanishing interaction between a D0 and an F-string.  The same discussion goes to the (F, D6)/D0 above and the end result is the same that there is no stringy interaction between a D0 and an F-string.  We therefore conclude that there is no stringy interaction between a D0 and an F-string. 
 
\section{The F/Dp case with $1 \le p \le 6$} 

We first consider the closed string tree cylinder amplitude for a system of a D3 and a D1, placed parallel at a separation $y$, with each carrying no flux. We denote this system as D1/D3. This amplitude can be computed using the D-brane boundary state representation\cite{Di Vecchia:1999fx}, following \cite{Jia:2019hbr}, to give
\be\label{D1D3-A}
\Gamma_{\rm{D1/D3}} = \frac{2\, V_{2}}{8 \pi^{2}\alpha'} \int_{0}^{\infty} \frac{d t}{t^{3}} \, e^{- \frac{y^{2}}{2\pi\alpha' t}} \prod_{n = 1}^{\infty} \frac{(1 + |z|^{4n})^{4}}{(1 - |z|^{4n})^{2} ( 1 - |z|^{2n})^{4}},
\ee
where $V_{2}$ is the volume of D1  worldvolume and $|z| = e^{- \pi t}$.  We now perform S-duality $g_{s} \to 1/g_{s}, \alpha' \to g_{s} \alpha'$ to the above amplitude to obtain the following for the system of an F-string and a D3 placed parallel at a separation $y$, denoting as F/D3, as
 \be\label{FD3-A}
 \Gamma_{\rm{F/D3}} = \frac{2\, V_{2}}{8 \pi^{2}\alpha' g_{s}} \int_{0}^{\infty} \frac{d t}{t^{3}} \, e^{- \frac{y^{2}}{2\pi\alpha' g_{s} t}}  \prod_{n = 1}^{\infty} \frac{(1 + |z|^{4n})^{4}}{(1 - |z|^{4n})^{2} ( 1 - |z|^{2n})^{4}}.
 \ee
  
 We now use T-duality acting on the amplitude (\ref{FD3-A}) either along a direction transverse or parallel (but not along the F-string direction) to D3. In general, the action of a T-duality along a direction transverse or parallel to a 1/2 BPS object in Type II theories can be summarized in Table 1
\begin{table}[htbp]\label{Table1}
\begin{center}  
 \begin{tabular}{|c|c|c|}
 \hline
& Parallel & Transverse \\
\hline
 Dp & D(p - 1) & D(p + 1) \\
 F & W & F \\
 W & F & W\\
 NS5 & NS5 & KK\\
 KK & KK & NS5\\
 \hline 
\end{tabular}
 \caption{The action of T-duality on a 1/2 BPS object in Type II theories.} 
 \end{center}
\end{table}
where ${\rm W}$ and ${\rm KK}$ denote waves and KK monopoles, respectively.    For concreteness, let us assume D3 to lie along $x^{1}, x^{2}, x^{3}$ with the F-string along $x^{1}$.  We denote the directions transverse to the D3 as $y_{m}$ with $m = 4, 5, \cdots 9$. The question is  how to implement the T-duality on the amplitude (\ref{FD3-A}).  Again for concreteness, we first choose to perform the T-duality along a direction transverse to the D3, for example, along $y_{4}$.  For this, we  need to compactify the $y_{4}$ to form a circle with a radius $a$ which can be equivalently implemented by placing a periodic array of D3 along the $y_{4}$ such that the k-th D3 is at $y^{(k)}_{4} = 2\pi a k$ with $k \in Z$, for example, following the prescription given in \cite{Lu:1999uca}.  We choose this D3 along the transverse space to be located at ${\bf y}^{(k)} = 2 \pi a k \, \hat e^{4}$ with $\hat e^{4}$ denoting the unit vector along the 4-th direction.  Suppose the  F-string is located at $\bf y$, then the interaction between this D3 and the F is, from (\ref{FD3-A}), as
\be\label{FD3n-A}
 \Gamma^{(k)}_{\rm{F/D3}} = \frac{2\, V_{2}}{8 \pi^{2}\alpha' g_{s}} \int_{0}^{\infty} \frac{d t}{t^{3}} \, e^{- \frac{y_{(k)}^{2}}{2\pi\alpha' g_{s} t}} \prod_{n = 1}^{\infty} \frac{(1 + |z|^{4n})^{4}}{(1 - |z|^{4n})^{2} ( 1 - |z|^{2n})^{4}},
 \ee 
 where $y_{(k)} = |{\bf y} - {\bf y}^{(k)}| = \sqrt{\bar y^{2} + (y_{4} - 2\pi a k)^{2}}$ with $\bar y^{2} = y^{2}_{5} + y^{2}_{6} + \cdots y^{2}_{9}$. Let us do the the following infinite sum 
 \be
 \sum_{k = - \infty}^{\infty} \Gamma^{(k)}_{\rm{F/D3}} =  \frac{2\, V_{2}}{8 \pi^{2}\alpha' g_{s}} \int_{0}^{\infty} \frac{d t}{t^{3}} \, \left( \sum_{k = - \infty}^{\infty} e^{- \frac{y_{(k)}^{2}}{2\pi\alpha' g_{s} t}}\right) \prod_{n = 1}^{\infty} \frac{(1 + |z|^{4n})^{4}}{(1 - |z|^{4n})^{2} ( 1 - |z|^{2n})^{4}}.
 \ee
 If we take a continuous limit, for example, by sending $a \to 0$, then the above infinite sum can be replaced by an integration
 \be
 \sum_{k = - \infty}^{\infty} = \frac{1}{2\pi a} \int_{- \infty}^{\infty} d z,
 \ee
 with $z = y_{4} - 2\pi k a$.  This process gives rise effectively to the compactification needed for the T-duality on the amplitude.  So we have now
 \be\label{FD4-A}
 \frac{1}{2\pi a} \int_{- \infty}^{\infty} dz\, \Gamma^{(z)}_{\rm{F/D3}} =  \frac{2^{\frac{1}{2}}\, V_{2}}{8 \pi^{2}\alpha' g^{\frac{1}{2}}_{s}} \frac{\sqrt{\alpha'}}{a} \int_{0}^{\infty} \frac{d t}{t^{\frac{5}{2}}} \, e^{- \frac{\bar y^{2}}{2\pi \alpha' g_{s} t}}  \prod_{n = 1}^{\infty} \frac{(1 + |z|^{4n})^{4}}{(1 - |z|^{4 n})^{2} ( 1 - |z|^{2n})^{4}},
 \ee
where we have used 
\be
 \frac{1}{2\pi a} \int_{- \infty}^{\infty} d z\, e^{- \frac{z^{2}}{2\pi\alpha' g_{s} t}} = \sqrt{\frac{g_{s}}{2}}\, \frac{\sqrt{\alpha'}}{a} \, t^{\frac{1}{2}}.
\ee 
If we make a choice of $a = \sqrt{\alpha'}$ for simplicity\footnote{With this choice, the T-duality $a \to \alpha'/a = a, \, g_{s} \to g_{s} \sqrt{\alpha'}/a = g_{s}$ and the parameters remain the same before and after the T-duality. In other words, we don't need to do anything about these parameters in performing the T duality.}, then the resulting amplitude in (\ref{FD4-A}) is nothing but the one for F/D4. It is
\be
\Gamma_{\rm{F/D4}} (y) =   \frac{2^{\frac{1}{2}}\, V_{2}}{8 \pi^{2}\alpha' g^{\frac{1}{2}}_{s}} \int_{0}^{\infty} \frac{d t}{t^{\frac{5}{2}}} \, e^{- \frac{y^{2}}{2\pi \alpha' g_{s} t}} \prod_{n = 1}^{\infty} \frac{(1 + |z|^{4n})^{4}}{(1 - |z|^{4n})^{2} ( 1 - |z|^{2n})^{4}} ,
\ee
where for the notation consistency, we use $y$ to replace $\bar y$ to denote the brane separation with the understanding now $y^{2} = y^{2}_{5} + \cdots y^{2}_{9}$.
 By repeating the above process along a direction transverse to the Dp brane at a time with $3 \le p \le 5$, we have in general
 \be\label{FDp-A}
 \Gamma_{\rm{F/Dp}} (y) =  \frac{2^{\frac{5 - p}{2}}\, V_{2}}{8 \pi^{2}\alpha' g^{\frac{5 - p}{2}}_{s}} \int_{0}^{\infty} \frac{d t}{t^{\frac{9 - p}{2}}} \, e^{- \frac{y^{2}}{2\pi \alpha' g_{s} t}}\prod_{n = 1}^{\infty} \frac{(1 + |z|^{4n})^{4}}{(1 - |z|^{4n})^{2} ( 1 - |z|^{2n})^{4}},
 \ee
 where $3 \le p \le 6$ and $|z| = e^{- \pi t}$.  Then how to obtain the cases of $p = 1$ and $p = 2$?  Given the above understanding of $3 \le p \le 6$, if we know the amplitude for an F-string and a D2, we can use the same procedure to obtain the known case of F/D3 as given in (\ref{FD3-A}). This must imply that the above formula should also work for the cases of $p = 1$ and $p = 2$. Or from the above procedure, we must infer that the dependence of the amplitude on the brane separation $y$ in the integration representation for an F-string and a D2  is also through the corresponding exponential factor $e^{- y^{2}/(2\pi \alpha' g_{s} t)}$. We then must end up with the above formulas working also for $p = 1$ and $p = 2$.   In other words, the amplitude (\ref{FDp-A}) is true for all $1 \le p \le 6$.  We now come to check that its long-range one for $1 \le p \le 6$ agrees with the result of (\ref{long-rangeF/Dp}).  For large $y$, we have from (\ref{FDp-A})
 \bea
 \Gamma_{\rm{F/Dp}} (y) &\approx&  \frac{2^{\frac{5 - p}{2}}\, V_{2}}{8 \pi^{2}\alpha' g^{\frac{5 - p}{2}}_{s}} \int_{0}^{\infty} \frac{d t}{t^{\frac{9 - p}{2}}} \, e^{- \frac{y^{2}}{2\pi \alpha' g_{s} t}}  =   \frac{2^{\frac{5 - p}{2}}\, V_{2}}{8 \pi^{2}\alpha' g^{\frac{5 - p}{2}}_{s}} \frac{(2 \pi \alpha' g_{s})^{\frac{7 - p}{2}}}{y^{7 - p}} \, \int_{0}^{\infty} d x \, x^{\frac{5 - p}{2}} e^{- x}\nn
 &=& \frac{V_{2} g_{s} \pi (2\pi \sqrt{\alpha'})^{5 - p}}{(7 - p) \Omega_{8 -p} \, y^{7 - p}},
  \eea
 which agrees perfectly with (\ref{long-rangeF/Dp}). In obtaining the last equality, we have used the gamma-function
 \be
 \Gamma \left(\frac{7 - p}{2}\right) = \int_{0}^{\infty} d x \, x^{\frac{5 - p}{2}} e^{- x},
 \ee 
 and the relation $\Gamma (1 + x) = x \Gamma(x)$ along the definition of $\Omega_{n}$ as given in the introduction. 
 
We could perform T-duality along $x^{1}$ direction, following Table 1, to end up with the amplitude $\Gamma_{\rm{W/D(p - 1)}}$ for a system of an W and a D(p -1).  However, this is not the focus of the present paper and we intend to discuss this case along with the more subtle amplitude $\Gamma_{\rm{Dp/KK}}$ for a Dp brane and an KK monopole, mentioned at the end of the following section,  in the near future. 
 
 \section{The Dp/NS5 case with $0 \le p \le 6$}
 For this case, we first compute the closed string tree cylinder amplitude, following  \cite{Jia:2019hbr}, for a system of a D3 and a D5 placed parallel at a separation $y$ with each carrying no worldvolume flux using their respective boundary state representation given in \cite{Di Vecchia:1999fx} as
\be\label{D3/D5-A}
\Gamma_{\rm{D3/D5}} = \frac{2\, V_{4} }{(8\pi^{2} \alpha')^{2}} \int_{0}^{\infty} \frac{d t}{ t^{2}} \, e^{- \frac{y^{2}}{2\pi \alpha' t}} \prod_{n = 1}^{\infty} \frac{(1 + |z|^{4n})^{4}}{(1 - |z|^{4n})^{2} (1 - |z|^{2n})^{4}},
\ee
 where $V_{4}$ is the volume of the D3 worldvolume and  $|z| = e^{- \pi t}$.  For concreteness, we take the D5 to be along $x^{1}, \cdots x^{5}$ directions and the D3 along $x^{1}, x^{2}, x^{3}$ directions but the two are localized at the directions transverse to both, for example the D5 at $\bf y = 0$ and the D3 at $ \bf y \neq 0$ with the separation $y^{2} = y^{2}_{6} + y^{2}_{7} + y^{2}_{8} + y^{2}_{9}$. Performing the IIB S-duality $\alpha' \to \alpha' g_{s}, g_{s} \to 1/g_{s}$ to the above amplitude, we end up with the amplitude for a system of a D3 and an NS5 as
 \be\label{D3NS5-A}
 \Gamma_{\rm{D3/NS5}} = \frac{2\, V_{4} }{(8\pi^{2} \alpha' g_{s})^{2}} \int_{0}^{\infty} \frac{d t}{ t^{2}} \, e^{- \frac{y^{2}}{2\pi \alpha' g_{s} t}} \prod_{n = 1}^{\infty} \frac{(1 + |z|^{4n})^{4}}{(1 - |z|^{4n})^{2} (1 - |z|^{2n})^{4}},
\ee
where the relevant quantities remain the same as before.  We now perform a T-duality along one longitudinal direction of the NS5 brane at a time such that the NS5 brane will not change according to Table 1.  Concretely, let us have this first along the $x^{4}$ direction. Given what we have done in the previous section, we have now the amplitude for a D4 and an NS5 with the same brane separation $y$ as
\bea\label{NS5D4-A}
\Gamma_{\rm{D4/NS5}} &=& \frac{1}{2\pi \sqrt{\alpha'}} \int_{ - \infty}^{\infty} d x^{4}\, \Gamma^{(x^{4})}_{\rm{D3/NS5}}\nn
&=& \frac{\hat V_{1}}{2\pi \sqrt{\alpha'}} \Gamma_{\rm{D3/NS5}}\nn
&=&  \frac{2\, V_{5} }{2\pi \sqrt{\alpha'} \, (8\pi^{2} \alpha' g_{s})^{2}} \int_{0}^{\infty} \frac{d t}{ t^{2}} \, e^{- \frac{y^{2}}{2\pi \alpha' g_{s} t}} \prod_{n = 1}^{\infty} \frac{(1 + |z|^{4n})^{4}}{(1 - |z|^{4n})^{2} (1 - |z|^{2n})^{4}},
\eea
where in the second equality we have used the fact that $\Gamma^{(x^{4})}_{\rm{D3/NS5}}$ is independent of $x^{4}$ and $\Gamma^{(x^{4})}_{\rm{D3/NS5}} = \Gamma_{\rm{D3/NS5}}$, and in the last equality we have set $V_{5} = V_{4} \hat V_{1}$ with $\hat V_{1} = \int_{-\infty}^{\infty} d x^{4}$ and $V_{5}$ the volume of the D4 worldvolume.   Following the same procedure by performing a T-duality along $x^{5}$, we end up with
\be
\Gamma_{\rm{D5/NS5}} = \frac{2\, V_{6} }{(2\pi \sqrt{\alpha'})^{2} \, (8\pi^{2} \alpha' g_{s})^{2}} \int_{0}^{\infty} \frac{d t}{ t^{2}} \, e^{- \frac{y^{2}}{2\pi \alpha' g_{s} t}} \prod_{n = 1}^{\infty} \frac{(1 + |z|^{4n})^{4}}{(1 - |z|^{4n})^{2} (1 - |z|^{2n})^{4}}, 
\ee
where $V_{6}$ is the volume of the D5 worldvolume.  If we perform T-duality along $x^{3}$, given the above, we should end up with
\be
\Gamma_{\rm{D2/NS5}} =  \frac{2\, V_{3} }{(2\pi \sqrt{\alpha'})^{-1} \, (8\pi^{2} \alpha' g_{s})^{2}} \int_{0}^{\infty} \frac{d t}{ t^{2}} \, e^{- \frac{y^{2}}{2\pi \alpha' g_{s} t}} \prod_{n = 1}^{\infty} \frac{(1 + |z|^{4n})^{4}}{(1 - |z|^{4n})^{2} (1 - |z|^{2n})^{4}}, 
\ee
 where $V_{3}$ is the volume of the D2 worldvolume.   Given the above, we have in general for $0 \le p \le 5$
 \be\label{DpNS5-A}
 \Gamma_{\rm{Dp/NS5}} = \frac{2\, V_{p + 1} }{(2\pi \sqrt{\alpha'})^{p - 3} \, (8\pi^{2} \alpha' g_{s})^{2}} \int_{0}^{\infty} \frac{d t}{ t^{2}} \, e^{- \frac{y^{2}}{2\pi \alpha' g_{s} t}} \prod_{n = 1}^{\infty} \frac{(1 + |z|^{4n})^{4}}{(1 - |z|^{4n})^{2} (1 - |z|^{2n})^{4}},
 \ee
 where $V_{p + 1}$ is the volume of the Dp worldvolume and the brane separation $y$ remains the same for all allowed $p$.  One can check that the large separation limit of (\ref{DpNS5-A}) gives
 \bea
 \Gamma_{\rm{DpNS5-A}} &\approx&  \frac{2\, V_{p + 1} }{(2\pi \sqrt{\alpha'})^{p - 3} \, (8\pi^{2} \alpha' g_{s})^{2}} \int_{0}^{\infty} \frac{d t}{ t^{2}} \, e^{- \frac{y^{2}}{2\pi \alpha' g_{s} t}}\nn 
 &=& \frac{V_{p + 1}\, (2\pi \sqrt{\alpha'})^{1 - p}}{4\pi\, g_{s}\, y^{2}},
 \eea
 where the second equality agrees perfectly with the formula (\ref{long-rangeDp/NS5}).
 
 As mentioned in the Introduction, the large separation interaction between an NS5 and a D6 placed parallel at a separation vanishes, independent of the separation $y$. This implies that the stringy one vanishes, too. This is also consistent with the vanishing interaction between an F-string and a D0 since the two are related to each other by the 10D Hodge duality. So we conclude that the stringy amplitude for a system of an NS5 and a D6 vanishes. It is clear that the interaction for any other system related to the D6/NS5 by T-dualities vanishes, too.
 
 We could also perform a T-duality on the amplitude $\Gamma_{\rm{Dp/NS5}}$ (\ref{DpNS5-A}) along a direction transverse to the NS5 brane to end up with the amplitude $\Gamma_{\rm{D(p + 1)/KK}}$ for  a system of a D(p + 1) and an KK monopole.  As always, this T-duality is more subtle and as mentioned at the end of the previous section we will come back to this one when we have a better understanding of this.  
 
\section{The amplitudes consistent with the IIB  S-duality}
The amplitudes obtained in the previous two sections appear good in IIA theory, but not all of them in IIB look to respect the IIB S-duality.  With the additional input of the IIB S-duality consistency, we will see that the amplitudes obtained in the previous two sections in both IIA and IIB  need modifications such that the resulting amplitudes are consistent with the IIB S-duality. For IIB theory, i.e., for  $p = 1, 3, 5$, we have from (\ref{FDp-A}) the following amplitudes
 \be\label{FDpIIB-A}
 \Gamma_{\rm{F/Dp}} (y) =  \frac{2^{\frac{5 - p}{2}}\, V_{2}}{8 \pi^{2}\alpha' g^{\frac{5 - p}{2}}_{s}} \int_{0}^{\infty} \frac{d t}{t^{\frac{9 - p}{2}}} \, e^{- \frac{y^{2}}{2\pi \alpha' g_{s} t}}\prod_{n = 1}^{\infty} \frac{(1 + |z|^{4n})^{4}}{(1 - |z|^{4n})^{2} ( 1 - |z|^{2n})^{4}},
 \ee
 and from (\ref{DpNS5-A}) the amplitudes 
  \be\label{DpNS5IIB-A}
 \Gamma_{\rm{Dp/NS5}} = \frac{2\, V_{p + 1} }{(2\pi \sqrt{\alpha'})^{p - 3} \, (8\pi^{2} \alpha' g_{s})^{2}} \int_{0}^{\infty} \frac{d t}{ t^{2}} \, e^{- \frac{y^{2}}{2\pi \alpha' g_{s} t}} \prod_{n = 1}^{\infty} \frac{(1 + |z|^{4n})^{4}}{(1 - |z|^{4n})^{2} (1 - |z|^{2n})^{4}}.
 \ee
 For either case with $p = 3$, there first appears  no issue since when we apply the S-duality $\alpha' \to \alpha' g_{s}, \, g_{s} \to 1/g_{s}$ to either amplitude, it gives the respective $\Gamma_{\rm{D1/D3}}$ as given in (\ref{FD3-A}) or the amplitude $\Gamma_{\rm{D3/D5}}$ as given in (\ref{D3/D5-A}).  However,  as we will see,  this is not the whole story when the IIB S-duality consistency is required  for each of  these amplitudes.  Let us first focus on the $p = 1$ and $p = 5$ in either case above for which we clearly do have issues.  
 
 From (\ref{FDpIIB-A}) for $p = 1$, we have
 \be
 \Gamma_{\rm{F/D1}} (y) =  \frac{ V_{2}}{2 \pi^{2}\alpha' g^{2}_{s}} \int_{0}^{\infty} \frac{d t}{t^{4}} \, e^{- \frac{y^{2}}{2\pi \alpha' g_{s} t}}\prod_{n = 1}^{\infty} \frac{(1 + |z|^{4n})^{4}}{(1 - |z|^{4n})^{2} ( 1 - |z|^{2n})^{4}}.
 \ee
 Applying the IIB S-duality, we have from the above
 \be
  \Gamma_{\rm{D1/F}} (y) =  \frac{ g_{s} \,V_{2}}{2 \pi^{2}\alpha' } \int_{0}^{\infty} \frac{d t}{t^{4}} \, e^{- \frac{y^{2}}{2\pi \alpha' t}}\prod_{n = 1}^{\infty} \frac{(1 + |z|^{4n})^{4}}{(1 - |z|^{4n})^{2} ( 1 - |z|^{2n})^{4}},
 \ee
 which is not as expected the same as the above $\Gamma_{\rm{F/D1}}$ though its large separation limit does give the same S-dual invariant one as that in (\ref{long-rangeF/Dp}) for $p = 1$.  To be consistent with the IIB S-duality, we propose that the underlying amplitude for this system is the following S-dual invariant one
 \bea\label{FD1IIB-A}
 \hat \Gamma_{\rm{F/D1}} (y) &\equiv&  \frac{1}{2} \left(\Gamma_{\rm{F/D1}} + \Gamma_{\rm{D1/F}}\right)\nn
 &=&  \frac{V_{2}}{4 \pi^{2}\alpha' g_{s}^{\frac{1}{2}} } \int_{0}^{\infty} \frac{d t}{t^{4}} \left( g^{-\frac{3}{2}}_{s} \, e^{- \frac{y^{2}}{2\pi \alpha' g_{s} t}} + g^{\frac{3}{2}}_{s}  e^{- \frac{y^{2}}{2\pi \alpha' t}}\right) \prod_{n = 1}^{\infty} \frac{(1 + |z|^{4n})^{4}}{(1 - |z|^{4n})^{2} ( 1 - |z|^{2n})^{4}},\nn
 \eea
 which is manifestly S-dual invariant and gives also the correct large separation limit as that in (\ref{long-rangeF/Dp}) for $p = 1$.  We now move to the $p = 5$ case from (\ref{DpNS5IIB-A}). The corresponding amplitude is
 \be
  \Gamma_{\rm{D5/NS5}} (y) = \frac{V_{6} }{2\pi^{2} \alpha' \, (8\pi^{2} \alpha' g_{s})^{2}} \int_{0}^{\infty} \frac{d t}{ t^{2}} \, e^{- \frac{y^{2}}{2\pi \alpha' g_{s} t}} \prod_{n = 1}^{\infty} \frac{(1 + |z|^{4n})^{4}}{(1 - |z|^{4n})^{2} (1 - |z|^{2n})^{4}}.
 \ee  
 Applying the IIB S-dual to this amplitude, we have
 \be
   \Gamma_{\rm{NS5/D5}} (y) = \frac{V_{6} }{2\pi^{2} \alpha' g_{s}\, (8\pi^{2} \alpha' )^{2}} \int_{0}^{\infty} \frac{d t}{ t^{2}} \, e^{- \frac{y^{2}}{2\pi \alpha' t}} \prod_{n = 1}^{\infty} \frac{(1 + |z|^{4n})^{4}}{(1 - |z|^{4n})^{2} (1 - |z|^{2n})^{4}},
 \ee   
  which once again is not  as expected the same as the above $\Gamma_{\rm{D5/NS5}}$ though again they have the same large separation S-dual invariant limit as that in (\ref{long-rangeDp/NS5}) for $p = 5$. By the same token, we propose the following S-dual invariant amplitude for this system as
  \bea\label{D5NS5IIB-A}
  \hat\Gamma_{\rm{D5/NS5}} (y) &\equiv& \frac{1}{2} \left(\Gamma_{\rm{D5/NS5}} + \Gamma_{\rm{NS5/D5}}\right)\nn
  &=& \frac{V_{6} }{4 \left(4 \pi^{2} \alpha' g^{\frac{1}{2}}_{s}\right)^{3}} \int_{0}^{\infty} \frac{d t}{ t^{2}} \left(\frac{ e^{- \frac{y^{2}}{2\pi \alpha' g_{s} t} } }{g_{s}^{\frac{1}{2}}} + g^{\frac{1}{2}}_{s}   e^{- \frac{y^{2}}{2\pi \alpha' t}}\right)\prod_{n = 1}^{\infty} \frac{(1 + |z|^{4n})^{4}}{(1 - |z|^{4n})^{2} (1 - |z|^{2n})^{4}},\nn
  \eea
which gives also the large separation limit (\ref{long-rangeDp/NS5}) for $p = 5$.   We now consider the $p = 5$ case in (\ref{FDpIIB-A}) and the amplitude is
 \be\label{FD5IIB-AB}
 \Gamma_{\rm{F/D5}} (y) =  \frac{V_{2}}{8 \pi^{2}\alpha'} \int_{0}^{\infty} \frac{d t}{t^{2}} \, e^{- \frac{y^{2}}{2\pi \alpha' g_{s} t}}\prod_{n = 1}^{\infty} \frac{(1 + |z|^{4n})^{4}}{(1 - |z|^{4n})^{2} ( 1 - |z|^{2n})^{4}},
 \ee
  which gives under the IIB S-duality
  \be
 \tilde \Gamma_{\rm{D1/NS5}} (y) =  \frac{V_{2}}{8 \pi^{2}\alpha' g_{s}} \int_{0}^{\infty} \frac{d t}{t^{2}} \, e^{- \frac{y^{2}}{2\pi \alpha'  t}}\prod_{n = 1}^{\infty} \frac{(1 + |z|^{4n})^{4}}{(1 - |z|^{4n})^{2} ( 1 - |z|^{2n})^{4}}.
 \ee  
 This one is however different from the amplitude $\Gamma_{\rm{D1/NS5}}$ obtained from (\ref{DpNS5IIB-A}) for $p = 1$ as 
 \be
 \Gamma_{\rm{D1/NS5}} (y) =  \frac{ V_{2} }{8\pi^{2} \alpha' g_{s}^{2}} \int_{0}^{\infty} \frac{d t}{ t^{2}} \, e^{- \frac{y^{2}}{2\pi \alpha' g_{s} t}} \prod_{n = 1}^{\infty} \frac{(1 + |z|^{4n})^{4}}{(1 - |z|^{4n})^{2} (1 - |z|^{2n})^{4}},
 \ee 
 though the two have the same large separation limit as given in (\ref{long-rangeDp/NS5}) for $p = 1$.   Applying the IIB S-duality to the above $\Gamma_{\rm{D1/NS5}}$, we end up with
 \be
 \tilde\Gamma_{\rm{F/D5}} (y) =  \frac{ V_{2} g_{s}}{8\pi^{2} \alpha' } \int_{0}^{\infty} \frac{d t}{ t^{2}} \, e^{- \frac{y^{2}}{2\pi \alpha'  t}} \prod_{n = 1}^{\infty} \frac{(1 + |z|^{4n})^{4}}{(1 - |z|^{4n})^{2} (1 - |z|^{2n})^{4}},
 \ee  
 which is also different from the one in (\ref{FD5IIB-AB}) though the two once again have the same large separation limit as given in (\ref{long-rangeF/Dp}) for $p = 5$.  By similar token, we now propose the amplitude being consistent with the IIB S-duality for the system of an F-string and a D5 brane as
 \bea\label{FD5IIB-A}
 \hat \Gamma_{\rm{F/D5}} (y) &\equiv& \frac{1}{2} \left(\Gamma_{\rm{F/D5}} + \tilde \Gamma_{\rm{F/D5}}\right) \nn
 &=&  \frac{V_{2}}{16 \pi^{2}\alpha' g_{s}^{\frac{1}{2}}} \int_{0}^{\infty} \frac{d t}{t^{2}} \left(g^{\frac{1}{2}}_{s}\, e^{- \frac{y^{2}}{2\pi \alpha' g_{s} t}}  + g^{\frac{3}{2}}_{s}\,e^{- \frac{y^{2}}{2\pi \alpha'  t}} \right) \prod_{n = 1}^{\infty} \frac{(1 + |z|^{4n})^{4}}{(1 - |z|^{4n})^{2} ( 1 - |z|^{2n})^{4}},\nn
 \eea
 while the amplitude being consistent with IIB S-duality for the system of a D1 and an NS5 is 
 \bea\label{D1NS5IIB-A}
 \hat \Gamma_{\rm{D1/NS5}} (y) &\equiv& \frac{1}{2}\left(\Gamma_{\rm{D1/NS5}} + \tilde \Gamma_{\rm{D1/NS5}}\right)\nn
 &=& \frac{ V_{2} }{16\pi^{2} \alpha' g_{s}^{\frac{1}{2}}} \int_{0}^{\infty} \frac{d t}{ t^{2}} \left(g^{-\frac{3}{2}}\, e^{- \frac{y^{2}}{2\pi \alpha' g_{s} t}} + g^{-\frac{1}{2}}_{s} \, e^{- \frac{y^{2}}{2\pi \alpha'  t}}\right)\prod_{n = 1}^{\infty} \frac{(1 + |z|^{4n})^{4}}{(1 - |z|^{4n})^{2} (1 - |z|^{2n})^{4}}.\nn
 \eea
 It is clear that the above $\hat \Gamma_{\rm{F/D5}}$ and $\hat\Gamma_{\rm{D1/NS5}}$ are now related to each other by the IIB S-duality and either gives its correct large separation limit as given in (\ref{long-rangeF/Dp}) for $p = 5$ for the former and in (\ref{long-rangeDp/NS5}) for $p = 1$ for the the latter. 
 
 Given the above $\hat \Gamma_{\rm{F/D1}}$ (\ref{FD1IIB-A}), we follow the procedure described in section 3 performing T-duality along a transverse direction at a time to end up with the general amplitude for $1\le p \le 6$ as
 \be\label{FDpAB-A}
 \hat \Gamma_{\rm{F/Dp}} = \frac{2^{\frac{5-p}{2}} \, V_{2}}{16 \pi^{2} \alpha' g_{s}^{\frac{1}{2}}} \int_{0}^{\infty} \frac{d t}{t^{\frac{9 - p}{2}}} \left(\frac{e^{- \frac{y^{2}}{2\pi \alpha' g_{s} t}}}{g^{\frac{4 - p}{2}}_{s}} +
 g^{\frac{3}{2}}_{s} \, e^{- \frac{y^{2}}{2\pi \alpha'  t}} \right)  \prod_{n = 1}^{\infty} \frac{(1 + |z|^{4n})^{4}}{(1 - |z|^{4n})^{2} (1 - |z|^{2n})^{4}},
 \ee 
where $y^{2} = y^{2}_{p + 1} + \cdots y^{2}_{9}$.  

Further given the amplitude $\hat \Gamma_{\rm{D5/NS5}}$ (\ref{D5NS5IIB-A}), we follow the procedure given in section 4 performing T-duality along a longitudinal direction of the NS5 at a time to end up with
the amplitude for $0 \le p \le 5$ as
\be\label{DpNS5AB-A} 
\hat \Gamma_{\rm{Dp/NS5}} =   \frac{V_{p + 1} }{4 \left(4 \pi^{2} \alpha' g^{\frac{1}{2}}_{s}\right)^{\frac{p + 1}{2}}} \int_{0}^{\infty} \frac{d t}{ t^{2}} \left(\frac{ e^{- \frac{y^{2}}{2\pi \alpha' g_{s} t} } }{g_{s}^{\frac{7 - p}{4}}} + \frac{ e^{- \frac{y^{2}}{2\pi \alpha' t}}}{g_{s}^{\frac{3 - p}{4}}}\right)\prod_{n = 1}^{\infty} \frac{(1 + |z|^{4n})^{4}}{(1 - |z|^{4n})^{2} (1 - |z|^{2n})^{4}},
\ee
where $y^{2} = y^{2}_{6} + \cdots + y^{2}_{9}$.

As non-trivial consistent checks, both (\ref{FDpAB-A}) and (\ref{DpNS5AB-A}) agrees perfectly with the corresponding (\ref{long-rangeF/Dp}) and (\ref{long-rangeDp/NS5}), respectively, when the respective large brane separation limit is taken. Further the amplitude (\ref{FDpAB-A}) for $p = 5$ agrees perfectly with that (\ref{FD5IIB-A})  while the amplitude (\ref{DpNS5AB-A}) for $p = 1$ agrees perfectly with (\ref{D1NS5IIB-A}).  In other words, the amplitude (\ref{FDpAB-A}) for $p = 5$ is perfectly S-dual to the one (\ref{DpNS5AB-A}) for $p = 1$.  This is a highly non-trivial check since both amplitudes start from their respective independent systems, for the former it is F/D1 while for the latter it is D5/NS5, then follow by the standard T-dualities.  The resulting system F/D5 for the former for $p = 5$ is indeed S-dual to the the resulting system D1/NS5 for the latter for $p = 1$.  The only requirement imposed is that both the amplitudes for F/D1 and for D5/NS5 are respective S-dual invariant, which has to be true, along with the reasonable assumption that T-dualities are valid non-perturbatively.  As stressed in the Introduction, the large separation leading amplitudes (\ref{long-rangeF/Dp}) and (\ref{long-rangeDp/NS5}) are both good for any string coupling $g_{s}$. We expect that the above (\ref{FDpAB-A}) and (\ref{DpNS5AB-A}) are also the leading non-perturbative stringy ones. Since the branes under consideration are still rigid, it is expected that the backreaction is not yet taken into account but the leading amplitudes are good for any string coupling.  This is unlike the amplitude (\ref{D1D3-A}) or (\ref{D3/D5-A}) which is good only for weak string coupling as mentioned earlier. 

Let us now take a close look at the amplitude (\ref{FDpAB-A}) or (\ref{DpNS5AB-A})  for $p = 3$.  For the former, we have
 \be\label{FD3B-A}
 \hat \Gamma_{\rm{F/D3}} = \frac{V_{2}}{8 \pi^{2} \alpha' g_{s}} \int_{0}^{\infty} \frac{d t}{t^{3}} \left(e^{- \frac{y^{2}}{2\pi \alpha' g_{s} t}} + g^{2}_{s} \, e^{- \frac{y^{2}}{2\pi \alpha'  t}} \right)  \prod_{n = 1}^{\infty} \frac{(1 + |z|^{4n})^{4}}{(1 - |z|^{4n})^{2} (1 - |z|^{2n})^{4}},
 \ee 
 while for the latter, we have
 \be\label{D3NS5B-A} 
\hat \Gamma_{\rm{D3/NS5}} =   \frac{V_{4} }{ \left(8 \pi^{2} \alpha' g_{s}\right)^{2}} \int_{0}^{\infty} \frac{d t}{ t^{2}} \left(e^{- \frac{y^{2}}{2\pi \alpha' g_{s} t}} + g_{s}\,e^{- \frac{y^{2}}{2\pi \alpha' t}}\right)\prod_{n = 1}^{\infty} \frac{(1 + |z|^{4n})^{4}}{(1 - |z|^{4n})^{2} (1 - |z|^{2n})^{4}},
\ee
either of which is different from the corresponding one (\ref{FD3-A}) or (\ref{D3NS5-A})  in that the first term in the bracket above counts only 1/2 of the corresponding one while the second term is new. Note that the two terms in the bracket give the equal large separation contribution and the total large separation limit gives the correct one (\ref{long-rangeF/Dp}) or (\ref{long-rangeDp/NS5}) for $p = 3$ as mentioned above.

We now comment on the above difference and have an understanding of this.  Note that the amplitude $\Gamma_{\rm{F/D3}}$ (\ref{FD3-A}) is obtained from the $\Gamma_{\rm{D1/D3}}$ (\ref{D1D3-A})  via the IIB S-duality.  As is clear and also noticed earlier, the two seemly appear to be consistent with the IIB S-duality. However, using the known D-brane technique, one important non-perturbative contribution to the amplitude has so far not been considered and this becomes manifest in the presence of the self-dual D3 brane.  This is precisely the source of non-consistency of the amplitudes  obtained in the previous two sections with the IIB S-duality. 

Note that in obtaining the $\Gamma_{\rm{D1/D3}}$  (\ref{D1D3-A}) using known D-brane technique, we need to keep the string coupling weak such that the D-branes considered can be taken as rigid to validate the computations.  This closed string tree cylinder amplitude counts only the contribution from the D3 emitting a virtual closed F-string, propagating for sometime and finally absorbed by the D1.  Let us give further close examination of the above amplitude $\hat\Gamma_{\rm{D1/D3}}$.  As mentioned above already, this self-dual D3 emits not only the virtual F-string but also the virtual closed D1 (or D-string).  This virtual closed D-string  gives the non-perturbative contribution to the second term in the bracket for the amplitude $\hat \Gamma_{\rm{D1/D3}}$ which can be obtained  from the amplitude $\hat \Gamma_{\rm{F/D3}}$ (\ref{FD3B-A}) via the IIB S-duality as
\be\label{D1D3B-A}
 \hat \Gamma_{\rm{D1/D3}} = \frac{V_{2}}{8 \pi^{2} \alpha' } \int_{0}^{\infty} \frac{d t}{t^{3}} \left(e^{- \frac{y^{2}}{2\pi \alpha'  t}} + g^{- 2}_{s} \, e^{- \frac{y^{2}}{2\pi \alpha' g_{s}  t}} \right)  \prod_{n = 1}^{\infty} \frac{(1 + |z|^{4n})^{4}}{(1 - |z|^{4n})^{2} (1 - |z|^{2n})^{4}}.
 \ee 
Note that the second term in the bracket above does appear as a non-perturbative correction with respect to the first term since it is inversely proportional to the string coupling square.  The first term comes from the exchange of the virtual closed  F-string while the second one comes from the exchange of the virtual closed D-string. Let us give an understanding of this second term contribution in the weak string coupling even though this amplitude is a leading non-perturbative one which is valid for any string coupling.  If we take the string coupling $g_{s} \to 0$, the first term is independent of $g_{s}$ and remains as it is while the second term, for any fixed $y$, has contribution only for large $t$ for which the infinite product can be taken as unity, i.e. the contribution from the excitations of the D-string can be ignored. The amplitude (\ref{D1D3B-A}) in the $g_{s} \to 0$ limit becomes
\be\label{D1D3B-wg}
 \hat \Gamma_{\rm{D1/D3}} = \frac{V_{2}}{8 \pi^{2} \alpha' } \int_{0}^{\infty} \frac{d t}{t^{3}} e^{- \frac{y^{2}}{2\pi \alpha'  t}} \,  \prod_{n = 1}^{\infty} \frac{(1 + |z|^{4n})^{4}}{(1 - |z|^{4n})^{2} (1 - |z|^{2n})^{4}} + \frac{\alpha' \, V_{2}}{2 \, y^{4}},
 \ee 
where the D-string contribution in the weak string coupling counts one half of the long-range interaction which can be obtained from \cite{Ouyang:2014bha} using the procedure described in the Introduction. This can be easily understood as follows. We know the mass spectrum for type IIB F-string is $M_{\rm{F}}^{2} = n /(4 \alpha')$ with $n = 0, 1, \cdots$ where $n = 0$ gives the massless spectrum. With this, the closed D-string spectrum is $M^{2}_{\rm{D1}} = n/(4 g_{s} \alpha')$ also with $n = 0, 1, \cdots$ where $n = 0$ gives the corresponding massless spectrum. In the weak coupling limit $g_{s} \to 0$, for the closed D-string, all except for the massless modes become superheavy and are too heavy to participate the exchange. In other words, only the massless modes contribute to the amplitude and this contribution is as expected the same as one half of the usual long-range one (Note that  the massless F-string modes contribute to the other half). 

For a general string coupling $g_{s}$,  the leading non-perturbative stringy amplitude is $\hat \Gamma_{\rm{D1/D3}}$ (\ref{D1D3B-A}) and the second term in the bracket is the non-perturbative contribution from the exchange of a virtual closed D-string. If we perform T-duality along or transverse to the brane system, we can end up with the resulting system exchanging not only the virtual closed F-string but also with the virtual closed D-brane.  If we begin with the open string one-loop annulus amplitude, which can be obtained using a Jacobi transformation to the above closed string tree cylinder one, the non-perturbative second term comes now from  a virtual open D-string with its two ends connecting the D1 and D3, respectively. If we also perform T-duality, the virtual open D-string will become higher-dimensional open D-brane if the T-duality is not along the original D-string direction. In either case, the resulting non-perturbative second term in the bracket in the amplitude gives the contribution from the exchange of the corresponding virtual closed D-brane or the virtual higher-dimensional open D-brane.  This non-perturbative contribution contains information about the coupling of the closed D-brane or the open D-brane \cite{Strominger:1995ac, Townsend:1996em}  and we will explore this further in a future work. Similar non-perturbative information about this coupling is also related to the first term in the bracket in the amplitude $\hat\Gamma_{\rm{F/D3}}$ (\ref{FD3B-A}) or in general in the amplitude $\hat\Gamma_{\rm{F/Dp}}$ (\ref{FDpAB-A}). 
  
 The same line of discussion goes also for the amplitude $\hat\Gamma_{\rm{D3/NS5}}$ (\ref{D3NS5B-A}) in replace of the one $\Gamma_{\rm{D3/NS5}}$ (\ref{D3NS5-A}) along with the issue with the amplitude  $\Gamma_{\rm{D3/D5}}$ (\ref{D3/D5-A}) computed using the usual D-brane technique in weak string coupling. We also expect that the important non-perturbative information can be extracted from the first term in the bracket in the amplitude $\hat\Gamma_{\rm{D3/NS5}}$ (\ref{D3NS5B-A}) and  in  the general amplitude $\hat \Gamma_{\rm{Dp/NS5}}$ (\ref{DpNS5AB-A}).
  
 In summary, if T-dualities are valid non-perturbatively and can be implemented as usual, consistency with the IIB S-duality gives the leading non-perturbative stringy amplitude for a system of an F-string and a Dp brane with $1 \le p \le 6$ as given in (\ref{FDpAB-A}) and the amplitude for a system of a Dp and an NS5 with $0 \le p \le 5$ as given in (\ref{DpNS5AB-A}).  The leading non-perturbative stringy amplitude vanishes for the system of an F-string and a D0 as well as for the system of a D6 and an NS5. The amplitudes so obtained modify the results from the usual standard D-brane technique which does not take into consideration of the non-perturbative contribution discussed above\footnote{The computation for the amplitude based on the usual D-brane technique overcounts twice the contribution from the exchange of F-string between two D branes with same or different dimensionality as indicated in the above discussion. This is due to the incorrect belief that the interaction is solely from the exchange of a virtual closed F-string between the two D branes (we now know that there is also a contribution from the virtual closed D-string or D-brane whose massless contribution is the same as that from the F-string). As such, the normalization used for the usual boundary state representation of D-brane is $\sqrt{2}$ too larger than its correct value. }.
 
 \section{Summary and conclusion}
Unlike the case for a system consisting of two D-branes placed parallel at a separation, there appears no basis to give a direct computation of the non-perturbative stringy amplitude for a system consisting of an NS brane and a D brane in a similar setting.  In this paper, we first make use of a particular system consisting of a D1 and a D3 placed parallel at a separation whose stringy amplitude can be computed directly using the known D brane technique. We then use the IIB S-duality and follow by various T-dualities to obtain the respective stringy amplitude for a system consisting of an F-string and a Dp brane with $1 \le p \le 6$. 
We argue that the stringy amplitude between an F-string and a D0 brane vanishes in general based on its large separation vanishing result and other known systems which involve the stringy interaction between F-strings and a D0 brane. 

By a similar token, we make use of the other particular system consisting of a D3 and a D5 placed parallel at a separation whose stringy amplitude can also be computed directly. With this, we use once again the IIB S-duality and follow by various T-dualities along the longitudinal directions of the resulting NS5 brane to obtain the respective stringy amplitude for a system consisting of a Dp brane and an NS5  brane with $0 \le p \le 5$.  For the system consisting of a D6 brane and an NS5 brane placed parallel at a separation, we use its large separation vanishing result and its 10 D Hodge dual relation to the vanishing result between a D0 brane and an F-string to conclude that this stringy amplitude vanishes also in general.

However, the above so obtained  stringy amplitudes in Type IIB string theory are not all of them to appear to be consistent with the underlying IIB S-duality.  The obvious ones are the $\Gamma_{\rm{F/D1}}$ and $\Gamma_{\rm{D5/NS5}}$ which are supposed to be the IIB S-dual invariant but they are not.  In addition, the so obtained $\Gamma_{\rm{F/D5}}$ from the original $\Gamma_{\rm{F/D3}}$ is not S-dual related to the $\Gamma_{\rm{D1/NS5}}$ obtained from the original $\Gamma_{\rm{D3/NS5}}$. We make further efforts in section 5 to resolve these inconsistencies  by proposing, in the case of inconsistency arising, the respective one being consistent with the underlying S-duality based on the obtained one and its S-dual one. The resulting ones are $\hat \Gamma_{\rm{F/D1}}$ and $\hat\Gamma_{\rm{D5/NS5}}$ along with the $\hat\Gamma_{\rm{F/D5}}$ and $\hat\Gamma_{\rm{D1/NS5}}$. These are not only being consistent with the IIB S-duality but also being consistent with T-dualities when we follow the procedure of performing various T-dualities as described in section 3 and 4. For example, when we perform T-dualities on the amplitude $\hat \Gamma_{\rm{F/D1}}$ along directions transverse to the underlying system to end up with the amplitude for the system of an F-string and a D5.  This amplitude turns out to be identical to the $\hat\Gamma_{\rm{F/D5}}$ obtained independently via an S-dual to the $\hat\Gamma_{\rm{D1/NS5}}$ which is obtained from $\hat \Gamma_{\rm{D3/NS5}}$ by T-dualities along the longitudinal directions to both D3 and NS5, a highly non-trivial consistent check.  

As discussed in rather detail in section 5,  assuming  that T-duality is valid non-perturbatively and can be implemented as usual, the consistency with the IIB S-duality gives the leading non-perturbative stringy amplitude for a system consisting of an F-string and a Dp brane with $1 \le p \le 6$ as given in (\ref{FDpAB-A}) and that for a system consisting of a Dp and an NS5 brane with $0 \le p \le 5$ as given in (\ref{DpNS5AB-A}).  For the former, this must imply that there is a modification as well as non-perturbative corrections to the usual amplitude computed using the standard D-brane technique for the system of a D1 and a D3.  For the latter, this happens also for the system of a D3 and D5.  This can be easily understood, as discussed in detail in section 5, from the D3 which emits not only the virtual closed F-string but also the virtual closed D-string for the underlying interaction amplitude. It is the virtual closed D-string that is usually not taken into consideration in the standard computation of the amplitude using the D-brane technique.  The combined applications of T-duality and S-duality remind us this important non-perturbative contribution to the amplitude.   The corresponding amplitude for the system of an F-string and a D0 or the system of a D6 and an NS5 remains zero.

Once again,  we would like to stress that all these leading non-perturbative stringy amplitudes, being consistent with T-dualities and the IIB S-dualities,  have their large separation limits to agree perfectly with the known results which can be computed using the low energy effective theories, i.e. the corresponding type II supergravities, plus the respective known couplings which can be read from the respective worldvolume effective theories.  This serves as an independent consistent check of the non-perturbative stringy amplitudes so obtained. 

As mentioned in section 5,  the leading non-perturbative stringy amplitude for either $\hat\Gamma_{\rm{D1/D3}}$ or $\hat\Gamma_{\rm{D3/D5}}$  or that obtained from the action of T-dualities on either of these two amplitudes contains the non-perturbative contribution which is due to the non-perturbative virtual D-brane (closed or open)  in the respective interaction.  Moreover there is also the non-perturbative contribution from the S-dual of either of these two amplitudes or that from the action of T-dualities on the S-dual amplitude.  These non-perturbative contributions provide us opportunities to explore the couplings of the underlying (closed or open) D branes with NS5 branes, F-strings and D branes.  Further these leading non-perturbative stringy amplitudes can also be used to determine the nature of the underlying interaction when more complicated systems such as non-threshold bound states involving F-strings or/and NS5 branes along with D branes are considered.   All this must teach us lesson and deepen our understanding of these couplings along with the underlying dynamics of these (closed or open) D branes.   It is clear that the leading non-perturbative stringy contribution to the amplitude is certainly important in revealing the underlying non-perturbative dynamics and may teach us even more lesson which cannot be obtained otherwise.    We wish to come to these along with some independent checks of the leading non-perturbative stringy amplitudes obtained in this paper in the near future.

% If in two-column mode, this environment will change to single-columnich 
% format so that long equations can be displayed. Use
% sparingly.
%\begin{widetext}
% put long equation here
%\end{widetext}

\section*{Acknowledgments}
The authors acknowledge the support by grants from the NNSF of China with Grant No: 12275264 and 12247103.


\begin{thebibliography}{99}

%\cite{Witten:1995im}
\bibitem{Witten:1995im}
E.~Witten,
``Bound states of strings and p-branes,''
Nucl. Phys. B \textbf{460}, 335-350 (1996)
doi:10.1016/0550-3213(95)00610-9
[arXiv:hep-th/9510135 [hep-th]].
%1629 citations counted in INSPIRE as of 19 Jul 2023

  %\cite{Breckenridge:1996tt}
\bibitem{Breckenridge:1996tt}
  J.~C.~Breckenridge, G.~Michaud and R.~C.~Myers,
  %``More D-brane bound states,''
  Phys.\ Rev.\  D {\bf 55}, 6438 (1997)
  [arXiv:hep-th/9611174].
  %%CITATION = PHRVA,D55,6438;%%

%\cite{Costa:1996zd}
\bibitem{Costa:1996zd}
  M.~S.~Costa and G.~Papadopoulos,
  ``Superstring dualities and p-brane bound states,''
  Nucl.\ Phys.\  B {\bf 510}, 217 (1998)
  [arXiv:hep-th/9612204].
  %%CITATION = NUPHA,B510,217;%%
  
  %\cite{Gava:1997jt}
\bibitem{Gava:1997jt}
E.~Gava, K.~S.~Narain and M.~H.~Sarmadi,
%``On the bound states of p-branes and (p+2)-branes,''
Nucl. Phys. B \textbf{504}, 214-238 (1997)
doi:10.1016/S0550-3213(97)00508-7
[arXiv:hep-th/9704006 [hep-th]].
%123 citations counted in INSPIRE as of 19 Jul 2023

%\cite{Di Vecchia:1997pr}
\bibitem{Di Vecchia:1997pr}
  P.~Di Vecchia, M.~Frau, I.~Pesando, S.~Sciuto, A.~Lerda and R.~Russo,
  ``Classical p-branes from boundary state,''
  Nucl.\ Phys.\  B {\bf 507}, 259 (1997)
  [arXiv:hep-th/9707068].
  
  %\cite{Sheikh-Jabbari:1997qke}
\bibitem{Sheikh-Jabbari:1997qke}
M.~M.~Sheikh-Jabbari,
``More on mixed boundary conditions and D-branes bound states,''
Phys. Lett. B \textbf{425}, 48-54 (1998)
doi:10.1016/S0370-2693(98)00199-3
[arXiv:hep-th/9712199 [hep-th]].
%67 citations counted in INSPIRE as of 19 Jul 2023

%\cite{Schwarz:1995dk}
\bibitem{Schwarz:1995dk}
J.~H.~Schwarz,
``An SL(2,Z) multiplet of type IIB superstrings,''
Phys. Lett. B \textbf{360}, 13-18 (1995)
[erratum: Phys. Lett. B \textbf{364}, 252 (1995)]
doi:10.1016/0370-2693(95)01405-5
[arXiv:hep-th/9508143 [hep-th]].

%\cite{Lu:1999uca}
\bibitem{Lu:1999uca}
J.~X.~Lu and S.~Roy,
``Nonthreshold (F, Dp) bound states,''
Nucl. Phys. B \textbf{560}, 181-206 (1999)
doi:10.1016/S0550-3213(99)00454-X
[arXiv:hep-th/9904129 [hep-th]].

%\cite{DiVecchia:1999uf}
\bibitem{DiVecchia:1999uf}
P.~Di Vecchia, M.~Frau, A.~Lerda and A.~Liccardo,
``(F,D(p)) bound states from the boundary state,''
Nucl. Phys. B \textbf{565}, 397-426 (2000)
doi:10.1016/S0550-3213(99)00632-X
[arXiv:hep-th/9906214 [hep-th]].

%\cite{Arfaei:1997hb}
\bibitem{Arfaei:1997hb}
H.~Arfaei and M.~M.~Sheikh Jabbari,
``Mixed boundary conditions and brane, string bound states,''
Nucl. Phys. B \textbf{526}, 278-294 (1998)
doi:10.1016/S0550-3213(98)00360-5
[arXiv:hep-th/9709054 [hep-th]].
%80 citations counted in INSPIRE as of 19 Jul 2023

%\cite{Ouyang:2014bha}
\bibitem{Ouyang:2014bha}
J.~Ouyang and C.~Wu,
``A classification of long-range interactions between two stacks of $p$ \textbackslash{}\& $p'$-branes,''
Commun. Theor. Phys. \textbf{63}, no.2, 195-208 (2015)
doi:10.1088/0253-6102/63/2/12
[arXiv:1409.0969 [hep-th]].
%2 citations counted in INSPIRE as of 28 Jul 2023

%\cite{Lu:2007kv}
\bibitem{Lu:2007kv}
J.~X.~Lu, B.~Ning, S.~Roy and S.~S.~Xu,
``On brane-antibrane forces,''
JHEP \textbf{08}, 042 (2007)
doi:10.1088/1126-6708/2007/08/042
[arXiv:0705.3709 [hep-th]].
%13 citations counted in INSPIRE as of 28 Jul 2023

%\cite{Strominger:1995ac}
\bibitem{Strominger:1995ac}
A.~Strominger,
``Open p-branes,''
Phys. Lett. B \textbf{383}, 44-47 (1996)
doi:10.1016/0370-2693(96)00712-5
[arXiv:hep-th/9512059 [hep-th]].
%704 citations counted in INSPIRE as of 22 Nov 2023

%\cite{Townsend:1996em}
\bibitem{Townsend:1996em}
P.~K.~Townsend,
``Brane surgery,''
Nucl. Phys. B Proc. Suppl. \textbf{58}, 163-175 (1997)
doi:10.1016/S0920-5632(97)00421-0
[arXiv:hep-th/9609217 [hep-th]].
%101 citations counted in INSPIRE as of 22 Nov 2023



%\cite{Lu:2023jxe}
\bibitem{Lu:2023jxe}
J.~X.~Lu,
``Understanding the open string pair production of the Dp/D0 system,'' JHEP \textbf{11}, 019(2023)
doi:10.1007/JHEP11(2023)019
[arXiv:2307.06594 [hep-th]] 
%1 citations counted in INSPIRE as of 07 Nov 2023


 %\cite{Di Vecchia:1999fx}
\bibitem{Di Vecchia:1999fx}
  P.~Di Vecchia and A.~Liccardo,
  ``D-branes in string theory. II,''
  arXiv:hep-th/9912275.
  %%CITATION = HEP-TH/9912275;%%
  
  %\cite{Jia:2019hbr}
\bibitem{Jia:2019hbr}
Q.~Jia, J.~X.~Lu, Z.~Wu and X.~Zhu,
``On D-brane interaction \textbackslash{}\& its related properties,''
Nucl. Phys. B \textbf{953}, 114947 (2020)
doi:10.1016/j.nuclphysb.2020.114947
[arXiv:1904.12480 [hep-th]].

  


%\cite{Porrati:1993qd}
%\bibitem{Porrati:1993qd}
  %M.~Porrati,
  %``Open strings in constant electric and magnetic fields,''
  % arXiv:hep-th/9309114.
  %%CITATION = HEP-TH/9309114;%%

  %%CITATION = NUPHA,B507,259;%%
    
  \end{thebibliography}
\end{document}